\documentclass[twocolumn,nofootinbib]{revtex4}

\usepackage{float}
\usepackage{graphicx}% Include figure files
\usepackage{dcolumn}% Align Tab.~columns on decimal point
\usepackage{bm}% bold math
\usepackage{amsmath}
\usepackage{hyperref}
\usepackage{xcolor}

\newcommand{\EDE}{\mathrm{EDE}}
\newcommand{\AIC}{\mathrm{AIC}}

\newcommand{\cdm}{\mathrm{cdm}}

\begin{document}

\title{Resolving the Hubble tension with Early Dark Energy}

\author{Laura Herold}
\email{lherold@mpa-garching.mpg.de}
\affiliation{Max Planck Institute for Astrophysics, Karl-Schwarzschild-Str. 1, 85748 Garching, Germany}
\author{Elisa G. M. Ferreira}
%\email{elisa.ferreira@ipmu.jp}
\affiliation{Kavli Institute for the Physics and Mathematics of the Universe (WPI), UTIAS, The University of Tokyo, Chiba 277-8583, Japan}
\affiliation{Instituto de F\'isica, Universidade de S\~ao Paulo - C.P. 66318, CEP: 05315-970, S\~ao Paulo, Brazil}

\begin{abstract}
Early dark energy (EDE) offers a solution to the so-called Hubble tension. Recently, it was shown that the constraints on EDE using Markov Chain Monte Carlo are affected by prior volume effects. The goal of this paper is to present 
%robust 
constraints on the fraction of EDE, $f_\EDE$, and the Hubble parameter, $H_0$, which are not subject to prior volume effects. We conduct a frequentist profile likelihood analysis considering \textit{Planck} cosmic microwave background, BOSS full-shape galaxy clustering, DES weak lensing, and SH0ES supernova data.
Contrary to previous findings,  we find that $H_0$ for the EDE model is in statistical agreement with the SH0ES direct measurement at $\leq 1.7\,\sigma$ for all data sets. For our baseline data set (\textit{Planck} + BOSS), we obtain $f_\EDE = 0.087\pm 0.037$ and $H_0 = 70.57 \pm 1.36\, \mathrm{km/s/Mpc}$ at $68\%$ confidence limit. We conclude that EDE is a viable solution to the Hubble tension. 
\end{abstract}

\maketitle

\section{Introduction}
\label{sec:intro}

The increasing precision of cosmological measurements  revealed a discrepancy known as the Hubble tension (see~\cite{Abdalla:2022yfr} for a review). 
The Hubble tension refers to the difference between direct measurements of $H_0$ and indirect measurements given a cosmological model. 
This tension reaches $5\,\sigma$ between the values obtained from the cosmic microwave background (CMB) data from \textit{Planck} for the $\Lambda$ Cold Dark Matter ($\Lambda$CDM) model~\citep{Planck_col_2020}, and from the Cepheid-calibrated Type Ia supernovae of the SH0ES project~\citep{Riess:2021jrx}.

While systematics are considered as a possible cause for the tension, growing interest has been given to the possibility that this tension points to new physics beyond the $\Lambda$CDM model. Among the most well studied proposed solutions to address this tension is the early dark energy (EDE) model~\citep{Poulin_2018, Poulin_2019, Smith_2020}, which introduces a new dark-energy component acting in the early universe.

This model was shown to successfully reduce the tension in $H_0$~\citep{Knox:2019rjx,Schoeneberg_2021_H0} when analyzed with \textit{Planck} CMB, Baryon Acoustic Oscillation, Pantheon supernova sample and data from SH0ES~\citep{Poulin_2018,Smith_2020}. Later it was pointed out in~\cite{Hill_2020,Ivanov_2020,DAmico_2021} that 
%when 
excluding the SH0ES measurement and including large-scale structure (LSS) probes like galaxy clustering and weak lensing leads to a tight upper limit on the amount of EDE, 
%was obtained, 
giving a value of $H_0$ compatible with the one from $\Lambda$CDM and not being able to solve the Hubble tension. 
Additionally, it was shown that the so-called $S_8$-tension, a tension in the amplitude of matter clustering, is worsened for the EDE model~\cite{Hill_2020,DAmico_2021,Secco:2022kqg}.

However, it was shown in~\cite{Herold:2021ksg}, previously hinted in~\cite{Niedermann_2020,Murgia:2020ryi,Smith:2020rxx} and later confirmed in~\cite{Gomez-Valent:2022hkb}, that the previous analyses of the EDE model using standard Bayesian Markov Chain Monte Carlo (MCMC) methods suffer from marginalization or prior volume effects that can bias the posteriors.

Prior volume effects are common effects in MCMC analyses that appear if the posterior is strongly influenced by the prior volume. In the case of the EDE model, the parameter structure of the model leads to large volume differences: When $f_\EDE$ approaches zero, the model reduces to $\Lambda$CDM; in this limit, the other parameters of the EDE model are unconstrained, which leads to an enhanced prior volume for $\Lambda$CDM and which can drive the posterior towards low fractions of EDE, $f_\EDE$, upon marginalization.

In view of these effects, it was suggested in~\cite{Herold:2021ksg} to use a frequentist profile likelihood. The profile likelihood and the Bayesian MCMC are complementary statistical tools since they address different statistical questions: While MCMC localizes large volumes in parameter space that fit the data well, the profile likelihood is based only on the minimum $\chi^2$, i.e. the best fit to the data, regardless of the size of the parameter volume. Therefore, the profile likelihood is reparametrization invariant \cite{hogg2013introduction} and, most importantly, is not influenced by prior volume effects. 

A profile likelihood of the EDE fraction, $f_{\mathrm{EDE}}$, resulted in a $f_{\mathrm{EDE}} = 0.072 \pm 0.036$~\cite{Herold:2021ksg} for \textit{Planck} data~\citep{Planck_col_2020} and Baryon Oscillation Spectroscopic Survey (BOSS) full-shape likelihood~\cite{Ivanov_2020_full-shape,DAmico_2020}, which is considerably higher than the MCMC result for the same data set. A similar analysis with free neutrino mass was performed in~\citep{Reeves:2022aoi}, with the goal of reducing $S_8$, finding a similar constraint (see~\cite{Hamann_2012,Planck_col_2014,Campeti_2022, CampetiKomatsu_2022} for application to other cases). 

The goal of this paper is to provide robust
%unbiased and conclusive 
constraints in the value of $H_0$ for the EDE model. We will assess the level of compatibility of the model-dependent $H_0$ constraints for the EDE model with the SH0ES direct measurement, revealing whether the EDE model can address the Hubble tension.

\section{Early Dark Energy}
\label{sec:EDE}

The EDE model contains a new component in the energy density of the universe that behaves like dark energy right after matter-radiation equality, but that dilutes away after recombination. The inclusion of this extra energy component decreases the sound horizon at the last scattering surface, which leads to an increase in $H_0$.

EDE~\citep{kamionkowski_2014,Karwal:2016vyq,Caldwell_2018} is the name given to a class of models satisfying the above dynamics (for some examples see~\cite{Reeves:2022aoi}). In this work, we use the canonical EDE model~\citep{Poulin_2019} which is described by a pseudoscalar field with the potential $V(\phi) = V_0 \, \left[ 1 - \cos (\phi/f) \right]^n$, where $V_0 = m^2 f^2$, $m$ and $f$ are the explicit and spontaneous symmetry breaking scales, respectively. Based on previous works~\citep{Poulin_2019,Smith_2020}, we study here the case of $n=3$, which satisfies the condition that the energy density of EDE dilutes faster than the one for matter.

One can relate the parameters of this model to the phenomenological parameters $f_\EDE$ and $z_c$, where $f_\EDE$ is the maximum fraction of EDE at the critical redshift $z_c$.  This field has a fixed initial value $\phi_i$, and becomes dynamical near $z_c$. These parameters together with the initial dimensionless value of the field $\theta_i \equiv \phi_i /f$, fully describe the EDE model. This phenomenological description is instrumental in making it clear that a higher $f_{\mathrm{EDE}}$ indicates a higher $H_0$; it was shown that $f_{\mathrm{EDE}} \sim 0.1$ is necessary to restore concordance in $H_0$ \citep{Bernal_2016,Knox:2019rjx}.

\section{Analysis Methods}
\label{sec:methodology}

\subsection{Data and modeling}
To model the EDE dynamics, we use the public \texttt{EDE\_CLASS\_PT} code~\citep{class_EDE_PT},
%\footnote{\url{https://github.com/Michalychforever/EDE_class_pt}}
an extension of the Einstein--Boltzmann solver \texttt{CLASS} \citep{Lesgourgues_2011, Blas_2011}, based on \texttt{CLASS\_EDE} \citep{Hill_2020} and \texttt{CLASS-PT} \citep{Chudaykin_2020}, a code based on the Effective Field Theory (EFT) of LSS \citep{BOSS_col_2017,Baumann_2012, Carrasco_2012} that allows to model the galaxy power spectrum up to mildly nonlinear scales. 

We consider the following data sets: \textit{Planck} 2018 TT, TE, EE, low$\ell$, lensing \citep{Planck_col_2020} (referred to as \textit{Planck}); the BOSS Data Release 12~\cite{BOSS_col_2017} full-shape power spectrum with a maximum wavenumber $k_\mathrm{max} = 0.25\,h$/Mpc using a consistent window-function normalization, which we implement along the lines of \citet{Beutler:2021eqq} and which corrects an inconsistency present before (referred to as BOSS); a Gaussian likelihood centered on the clustering amplitude of matter, $S_8=\sigma_8\sqrt{\Omega_m/0.3} = 0.776 \pm 0.017$, measured by the Dark Energy Survey Year 3 analysis (referred to as DES)~\cite{Abbott_2022}\footnote{Using a Gaussian likelihood is an approximation but it was tested in~\cite{Hill_2020} for DES Y1 that the difference to the full likelihood is small for the EDE model.}; and a Gaussian likelihood centered on $H_0 = 73.04 \pm 1.04$ measured by SH0ES~\citep{Riess:2021jrx} (referred to as SH0ES). 

We sample the $\Lambda$CDM parameters \{$\omega_b$, $\omega_\cdm$, $\theta_s$, $A_s$, $n_s$, $\tau_\mathrm{reio}$\}, the EDE parameters \{$f_\EDE$, $\log(z_c)$, $\theta_i$\}, along with the \textit{Planck} and EFT nuisance parameters. Following the convention of the \textit{Planck} collaboration \citep{Planck_col_2020}, we model the neutrino sector by two massless and one massive neutrino species with $m_\nu = 0.06\,$eV.

\subsection{Statistical inference: MCMC and profile likelihood}

We perform both a Bayesian MCMC and a frequentist profile likelihood analysis using \texttt{MontePython} \citep{Brinckmann_2018} with the Metropolis--Hastings algorithm \citep{Metropolis_1953, Hastings_1970}. We assume 
%flat priors on all $\Lambda$CDM parameters, 
the same priors as~\cite{Philcox_2022} on the EFT nuisance parameters, and the same priors as~\cite{Hill_2020} on the EDE parameters. We require the Gelman-Rubin convergence criterion $R-1 < 0.05$. 

Following the methodology in our previous works~\cite{Herold:2021ksg, Reeves:2022aoi}, we construct a profile likelihood by fixing the parameter of interest to different values and minimizing $\chi^2=-2\ln {\mathcal L}$ with respect to all other parameters of the model, where $\mathcal{L}$ denotes the likelihood. The $\Delta\chi^2$ as a function of the parameter of interest is the profile likelihood. For the minimization, we adopt a simulated annealing approach based on the method used by \citet{Schoeneberg_2021_H0} (see also~\cite{Hannestad_1999}). 
As in our previous work~\cite{Herold:2021ksg}, we construct a confidence interval from the profile likelihood following the prescription by \citet{Feldman_1998}, which extends the procedure by \citet{Neyman:1937uhy} and is also valid at a physical boundary. We quote confidence intervals obtained from profile likelihoods (MCMC) as bestfit (mean) $\pm 1\,\sigma$.  

\section{Results and discussion}
\label{sec:results}

\begin{figure}[t!]
    \centering
    \includegraphics[scale=0.5]{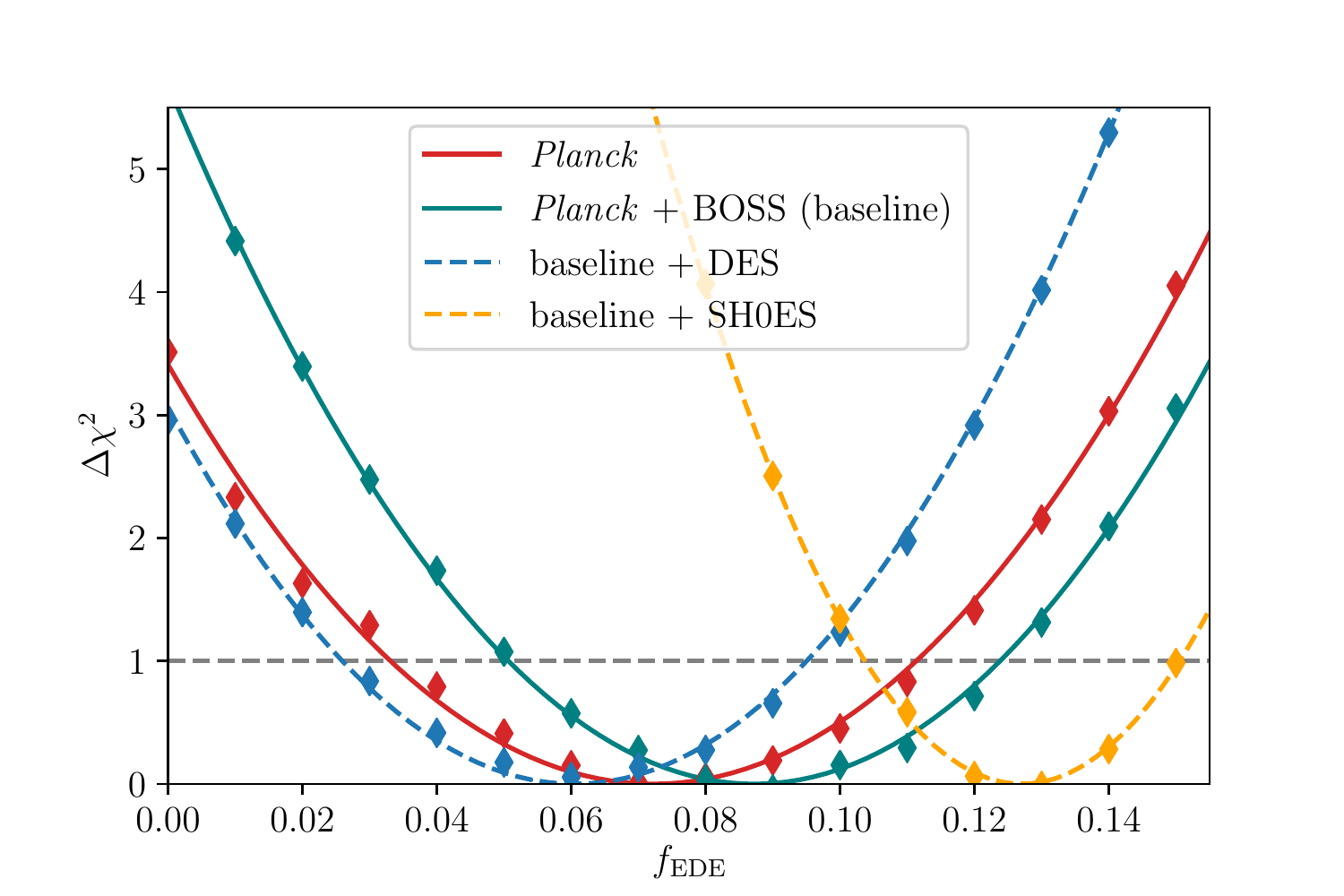}
     \caption{\label{fig:fEDE_all} Profile likelihoods (markers) for the maximum fraction of EDE, $f_\EDE$, for different data sets. The intersection of the parabola fit (lines) with $\Delta\chi^2=1$ (horizontal dashed line) gives the $1\,\sigma$ confidence interval in the approximate Neyman construction.}
\end{figure}

\begin{figure}[t!]
    \centering
    \includegraphics[scale=0.5]{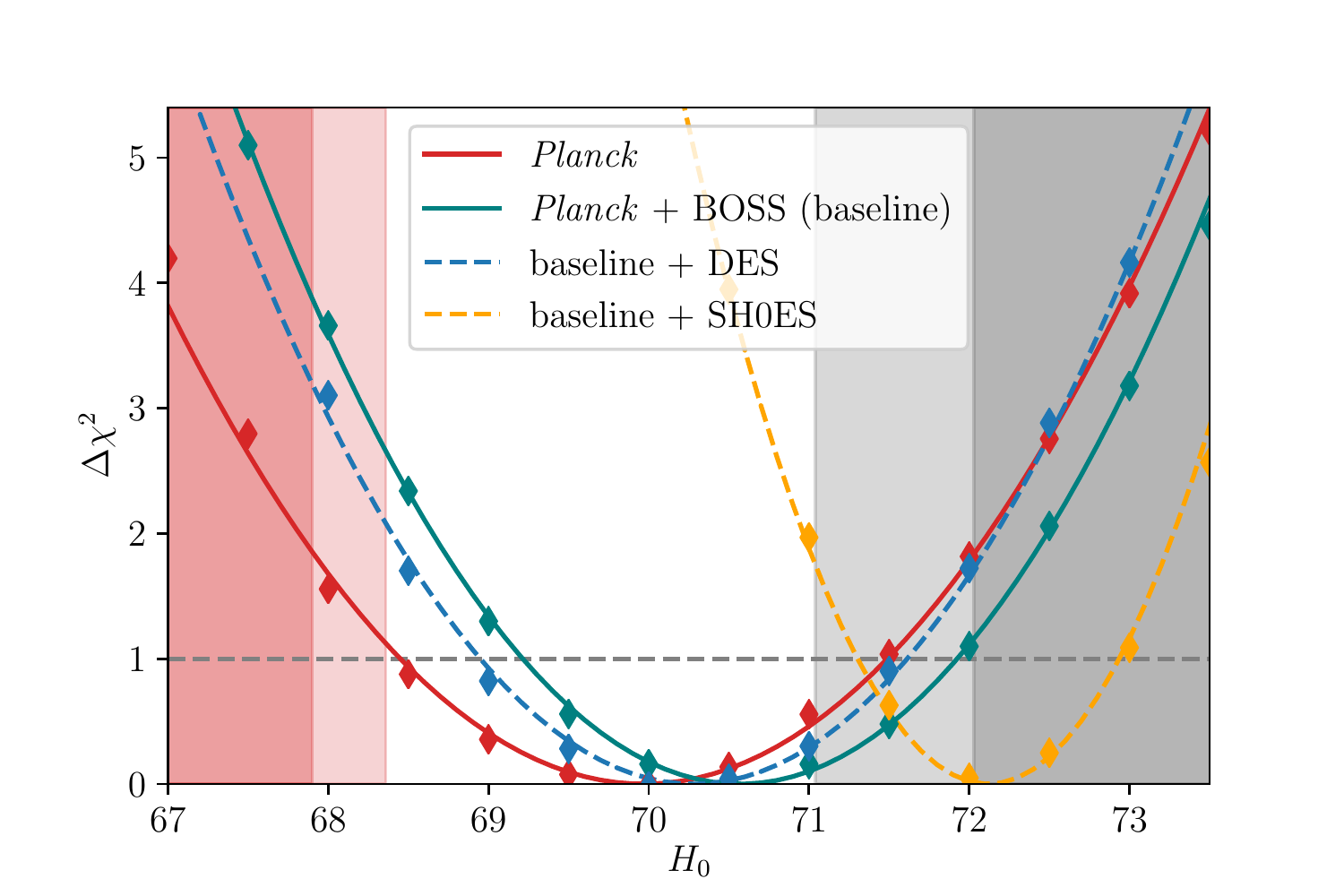}
     \caption{\label{fig:H0_all} Profile likelihoods for the Hubble parameter, $H_0$, for different data sets. The red vertical region corresponds to the $1\sigma$ and $2\sigma$ contours for $H_0$ from \textit{Planck} 2018 for $\Lambda$CDM, while the grey region corresponds to the $1\,\sigma$ and $2\,\sigma$ contours for the direct measurement by SH0ES.}
\end{figure}

Fig.~\ref{fig:fEDE_all} and Fig. \ref{fig:H0_all} present the final result of our profile likelihood analysis for $f_\EDE$ and $H_0$ for different datasets, with final confidence intervals summarized in Fig.~\ref{fig:summary} and Table~\ref{tab:model_comparison}.

\begin{table*}
    \centering
    \begin{tabular}{|l|c|c|c|c||c|c|}
    \hline
    Data set     & $\chi^2(\Lambda$CDM)  & $\chi^2$(EDE)  & $\Delta\chi^2$ & $\Delta$\,AIC &$f_\EDE$ & $H_0$  (consistency w.\ SH0ES) \\
    \hline
    \textit{Planck}                  & 2774.24   & 2770.72   & $-3.52$ & $+2.48$ & $0.072\pm0.039$           & $69.97\pm1.52$ ($1.7\,\sigma$)\\
    \textit{Planck}+BOSS (base)      & 3045.65   & 3039.98   & $-5.67$ & $+0.33$ & $0.087\pm0.037$           & $70.57\pm1.36$ ($1.4\,\sigma$)\\
    Baseline + DES           & 3052.06   & 3049.13   & $-2.93$ & $+3.07$ & $0.061^{+0.035}_{-0.034}$ & $70.28\pm1.33$ ($1.6\,\sigma$)\\
    Baseline + SH0ES           & 3068.44   & 3042.08   &$-26.36$ &$-20.36$ & $0.127\pm0.023$           & $72.12\pm0.82$  ($0.69\,\sigma$)            \\
    \hline
    \end{tabular}
    \caption{The $\chi^2$ values of the $\Lambda$CDM and bestfit EDE models, the difference $\Delta\chi^2 = \chi^2(\mathrm{EDE}) - \chi^2(\Lambda\mathrm{CDM})$, the Akaike information criterion (AIC), the constraints on $f_\EDE$ and $H_0$, and the compatibility with the SH0ES measurement in units of $\sigma$ for the different data sets considered in this work.}
     \label{tab:model_comparison}
\end{table*}

\subsection{\textit{Planck} + BOSS full-shape analysis (baseline)}
\label{sec:baseline}

Our baseline data set consists of \textit{Planck} CMB and BOSS galaxy clustering data (solid teal lines in Figs. \ref{fig:fEDE_all},~\ref{fig:H0_all}). The confidence intervals obtained from the profile likelihood are:
\begin{equation}
    f_\EDE = 0.087\pm 0.037,\ 
    H_0 = 70.57 \pm 1.36\, \mathrm{km/s/Mpc}\,.
\end{equation}
To assess parameter consistency, we report the one-dimensional difference between the bestfits of the two measurements divided by the quadrature sum of the $1\,\sigma$ errors.
%~\citep{Smith:2022iax}. 
We find that $H_0$ obtained from the baseline data set within the EDE model is consistent with SH0ES at $1.4\,\sigma$.
%~\footnote{Assuming a Gaussian around the bestfit values is a justified assumption for the result from the PL given its parabolic shape. In general, this assumption might overestimate tensions~\citep{Smith:2022iax}} 

Compared to $\Lambda$CDM, the goodness of fit to the data improves by $\Delta\chi^2=-5.67$ for the EDE model with $f_\EDE = 0.09$ (see Table~\ref{tab:model_comparison})\footnote{We cite $\chi^2$ and bestfit parameters for the EDE cosmology with fixed $f_\EDE$ that is closest to the global minimum (minimum of the profile likelihood). The error of this approximation is negligible compared to the $1\,\sigma$ statistical uncertainty and can only lead to an \textit{underestimation} of the improvement of fit for EDE.}. 
To assess whether the data prefers EDE with extra parameters over $\Lambda$CDM, we compute the Akaike information criterion (AIC)~\citep{burnham2002model}, which penalizes additional parameters and is defined as $\Delta\AIC = \Delta \chi^2 + 2\,\Delta N$, where $\Delta N$ is the number of additional parameters of the extended model (for EDE: $\Delta N=3$). We find $\Delta\AIC = +0.33$, i.e. a not statistically significant preference for $\Lambda$CDM over EDE.

For direct comparison, we run an MCMC analysis for the same data set and find a tight upper limit $f_\EDE < 0.072$ (at $95\%$ confidence), and $H_0 = 68.55_{-1.06}^{+0.62}\, \mathrm{km/s/Mpc}$, which is in tension with SH0ES at $3.7\,\sigma$. 
%The constraints on both parameters are considerably smaller than in the profile likelihood analysis. 
As pointed out previously~\cite{Herold:2021ksg}, the difference to the profile likelihood result can be explained by prior volume effects affecting the results of the MCMC results.

The constraints on $f_\EDE$ and $H_0$ found here are slightly higher than those from a profile likelihood analysis with the previously widely used BOSS likelihood using an inconsistent normalization ($f_\EDE = 0.072 \pm 0.036$ \cite{Herold:2021ksg}). The consistent window-function normalization leads to higher values of $S_8$. Since $S_8$ is increased in EDE cosmologies compared to $\Lambda$CDM, a higher $S_8$ allows for more EDE. This is in agreement with \citet{Simon_2022ede}, who use MCMC to constrain EDE and find a weaker upper limit on $f_\EDE$ with the consistent window-function normalization as compared to the inconsistent normalization.

With the profile likelihood analysis, we also find shifts in other cosmological parameters compared to $\Lambda$CDM: the bestfit $n_s$ increases from $0.968$ ($\Lambda$CDM) to $0.983$ (bestfit EDE cosmology, $f_\EDE = 0.09$), and $\omega_\cdm$ from $0.120$ ($\Lambda$CDM) to $0.129$ ($f_\EDE = 0.09$), which can be understood as a compensation of a suppressed early Sachs-Wolfe effect in EDE cosmologies~\cite{Vagnozzi_2021}. The most notable change is in $S_8$, which increases from $0.828$ ($\Lambda$CDM) to $0.840$ ($f_\EDE = 0.09$), worsening the so-called $S_8$-tension with weak-lensing experiments~\citep{DiValentino:2018gcu,Nunes:2021ipq}.

\subsection{Baseline + DES}

Since EDE cosmologies feature higher $S_8$~\cite{Smith_2020, Hill_2020, Secco:2022kqg}, including weak lensing measurements into the analysis is an important test for EDE. In this section, we include a Gaussian likelihood from DES\footnote{We did not include likelihoods for HSC~\cite{Hikage_2019} and KiDS~\cite{Asgari_2021} simultaneously since there is non-negligible cross-correlation between the data sets. Using a combined weak-lensing likelihood would be an important further check.} with $S_8=0.776 \pm 0.017$ along with the baseline data set (blue dashed lines in Figs.~\ref{fig:fEDE_all},~\ref{fig:H0_all}). The profile likelihood analysis yields: 
\begin{equation}
    f_\EDE = 0.061^{+0.035}_{-0.034},\ H_0 = 70.28 \pm 1.33\, \mathrm{km/s/Mpc}\,.
\end{equation}
As expected, we find smaller $f_\EDE$ and $H_0$ than those from the baseline data set, but  $H_0$ is still consistent with SH0ES at $1.6\,\sigma$. The improvement of the fit compared to $\Lambda$CDM, $\Delta\chi^2 = -2.93$, is smaller than for the baseline result. The worsening can be attributed mainly to the contribution from the $S_8$ likelihood. The bestfit $S_8$ for $\Lambda$CDM, $S_8=0.812$, and the bestfit EDE model $f_\EDE = 0.06$, $S_8=0.817$, are comparable but both are higher than the DES measurement, $S_8=0.776$. The AIC shows a mild preference for $\Lambda$CDM over EDE, $\Delta\AIC = +3.07$.

The trend of a decreasing $f_\EDE$ and $H_0$ when including an $S_8$ likelihood is similar as in previous MCMC analyses~\cite{Smith_2020, Hill_2020,Secco:2022kqg} but the effect in the profile likelihood is less pronounced since it is not overlaid by prior volume effects. While the MCMC results suggest that EDE is not able to solve the $H_0$ tension, the profile-likelihood result for $H_0$ from the baseline + DES data set is in statistical agreement with the SH0ES measurement.

\subsection{Baseline + SH0ES}

Given that the value of $H_0$ for the EDE baseline data set is consistent with the SH0ES measurement at $1.4\,\sigma$, it is sensible to combine both data sets. A profile-likelihood analysis of the baseline data set with a Gaussian likelihood centered on the measurement by the SH0ES experiment, $H_0 = 73.04 \pm 1.04$ (yellow dashed lines in Figs.~ \ref{fig:fEDE_all},~\ref{fig:H0_all}) yields:
\begin{equation}
    f_\EDE = 0.127 \pm 0.023,\ 
    H_0 = 72.12 \pm 0.82\, \mathrm{km/s/Mpc}. 
\end{equation}
This constraint of $H_0$ is consistent with SH0ES at $0.69\,\sigma$. We find an improvement of fit of the EDE model compared to $\Lambda$CDM by $\Delta\chi^2=-26.36$, where the main contribution to the $\Delta\chi^2$ comes from the SH0ES-$H_0$ likelihood, $\Delta\chi^2_\mathrm{SH0ES}=-18.47$. The AIC shows a strong preference for the EDE model over $\Lambda$CDM, $\Delta\AIC = -20.36$. 
The profile likelihood constraints are consistent with previous MCMC constraints including SH0ES data~\citep{Smith_2020,Hill_2020, DAmico_2021,Murgia:2020ryi}\footnote{With the exception of the result from~\citet{DAmico_2021} for Planck+BAO+SnIa(Pantheon)+BOSS full-shape power spectrum+SH0ES, which is consistent with our result at $\sim2\,\sigma$.} at $< 1\,\sigma$.

The constraints of $H_0$ and $f_\EDE$ within the EDE model for the baseline + SH0ES data set are consistent with the constraints for all other data sets considered here at  $< 1.3\,\sigma$ and $< 1.6\,\sigma$, respectively.

\subsection{\textit{Planck}-only constraint and comparison to ACT}

Lastly, we probe the constraining power of the \textit{Planck} CMB data alone. We find
\begin{equation}
    f_\EDE = 0.072 \pm 0.039,\ 
    H_0 = 69.97 \pm 1.52\,\mathrm{km/s/Mpc}.
\end{equation}
The $H_0$ constraint is consistent with SH0ES at $1.7\,\sigma$. We find an improvement of fit of $\Delta\chi^2 = -3.52$. This improvement is dominated by the \textit{Planck} high-$\ell$ likelihood with $\Delta\chi^2_{\mathrm{high-}\ell} = -2.90$. The AIC shows a mild preference of $\Lambda$CDM over EDE, $\Delta\AIC=+2.48$. 

The relatively high $f_\EDE$ preferred by \textit{Planck} in the profile likelihood analysis is interesting in light of the preference for $f_\EDE$ in an MCMC analysis of Atacama Cosmology Telescope (ACT) CMB data~\cite{Choi_2020}. The profile likelihood constraints of $f_\EDE$ from \textit{Planck} are consistent at $<1\,\sigma$ with MCMC constraints from ACT ($f_\EDE = 0.091^{+0.020}_{-0.036}$ for the baseline data set in~\cite{Hill_2021_ACT}, see also~\cite{Smith:2022hwi, LaPosta_2021}). 
The difference between the results from \textit{Planck} and ACT from MCMC analyses is likely due to prior volume effects in the MCMC analysis for \textit{Planck}. 
The strong preference for the EDE model over $\Lambda$CDM that was found for ACT~\cite{Hill_2021_ACT} seems to indicate that the constraints from this data set are less affected by prior volume effects.

\section{Conclusion}
\label{sec:conclusions}

\begin{figure}
    \centering
    \includegraphics[scale=0.5]{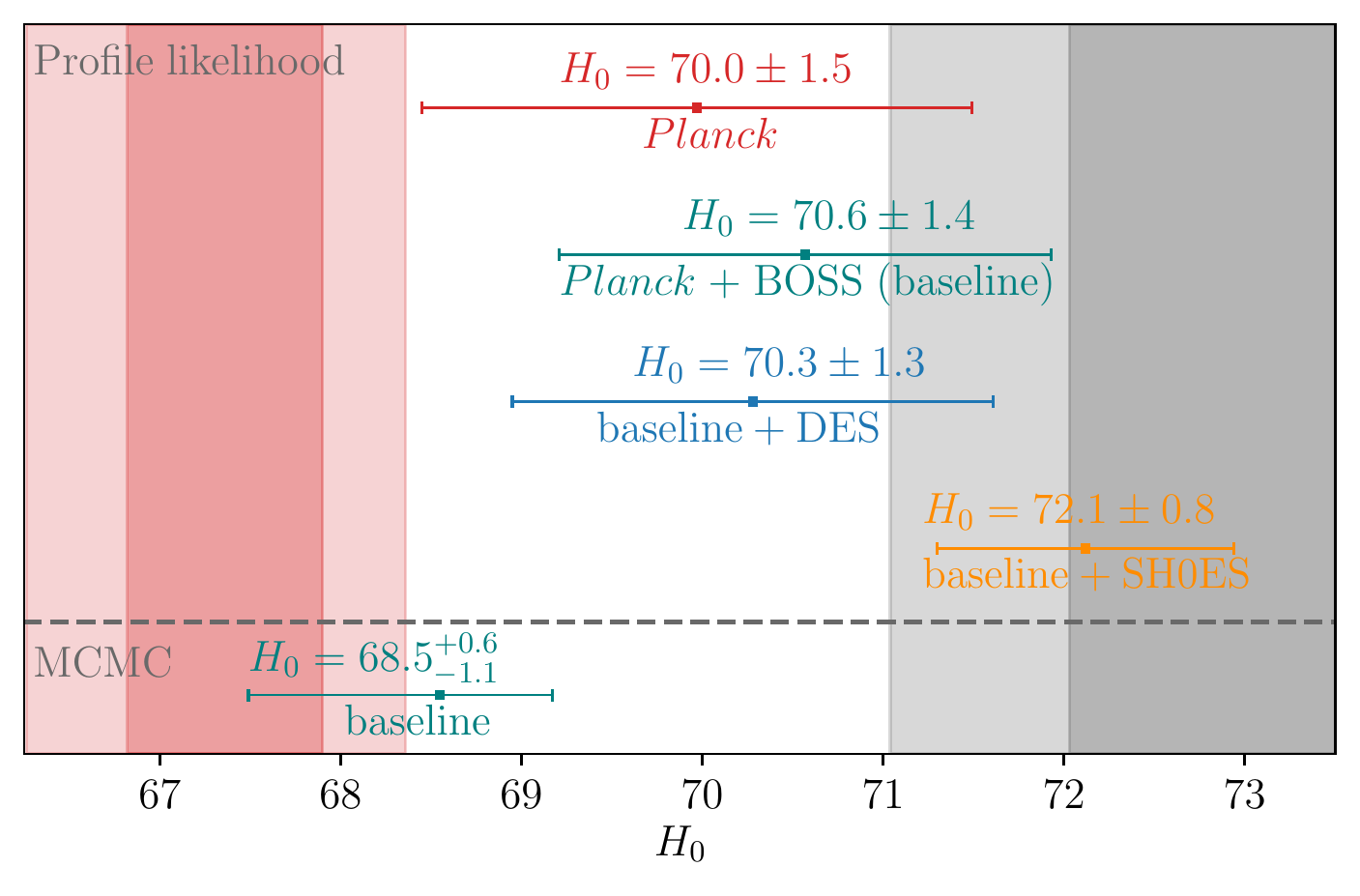}
     \caption{\label{fig:summary} Constraints of $H_0$ within the EDE model for different data sets. The top four errorbars show constraints from the profile likelihood, whereas the bottom errorbar shows the constraint from MCMC. For comparison, the red shaded area corresponds to the $1\,\sigma$ and $2\,\sigma$ constraint from \textit{Planck}~\cite{Planck_col_2020} assuming $\Lambda$CDM and the grey shaded area to the $1\,\sigma$ and $2\,\sigma$ constraint from SH0ES~\cite{Riess:2021jrx}.}
\end{figure}

In this paper, we obtained constraints on the value of $H_0$ for the EDE model, which are not subject to prior volume effects, using a frequentist profile likelihood and assessed the viability of EDE as a solution to the Hubble tension.  

It was previously concluded from MCMC analyses that EDE is not able to resolve the $H_0$ tension and simultaneously fit different cosmological data. We find a similar result from the MCMC analysis of our baseline data set (bottom errorbar in Fig.~\ref{fig:summary}).
As was previously shown in~\cite{Herold:2021ksg}, MCMC analyses of the EDE model are affected by marginalization or prior volume effects.
Therefore, we used the profile likelihood to obtain confidence intervals for $H_0$ (Fig~\ref{fig:summary}) and to assess consistency with other measurements and the resolution of the tension.

We assessed whether the data prefers EDE over $\Lambda$CDM using the AIC, which takes into account that the EDE model has three additional parameters compared to $\Lambda$CDM. The AIC shows a mild preference for $\Lambda$CDM for the baseline data set, the baseline + DES and the Planck-only data sets.  Only when adding SH0ES, there is a clear preference for the EDE model over $\Lambda$CDM. Therefore, EDE presents a good fit to CMB and LSS even when penalizing the additional parameters of EDE. 
 
Our baseline data set yields $H_0=70.57 \pm1.36\,$km/s/Mpc, which is consistent with SH0ES at $1.4\, \sigma$. This value is considerably higher than the MCMC result, reinforcing the evidence for prior volume effects in the Bayesian analysis.

Adding a likelihood centered on the $S_8$ measurement from DES decreases $f_{\mathrm{EDE}}$ with respect to the baseline data set, translating into a mild decrease in $H_0$. This is expected since EDE cosmologies show a positive correlation of $S_8$ with $f_{\mathrm{EDE}}$ and $H_0$~\cite{Secco:2022kqg}. However, this decrease is much smaller than the one found in previous MCMC analysis. The $H_0$ for baseline + DES is consistent with the SH0ES value at $1.6\,\sigma$. Hence, even for the most constraining data combination for EDE considered here, we find an agreement with SH0ES.

Given that the value of $H_0$ for the baseline data set is consistent with the SH0ES measurement, we can combine both data sets.  As expected from previous analyses, including SH0ES to the baseline data set results in an even higher $H_0$ than for the baseline data set. This is consistent with the SH0ES measurement at $0.69\,\sigma$. 

Finally, we find that the $H_0$ constraint from \textit{Planck} data alone is compatible with SH0ES, and interestingly also in agreement with previous works performing an MCMC analysis with ACT data. Considering the relative $\chi^2$ contributions for all likelihoods considered in this work, we find that (apart from SH0ES), the \textit{Planck} high-$\ell$ likelihood dominates the improvement of fit compared to all other data sets. 

For all data combinations, the $H_0$ value obtained with the profile likelihood analysis is consistent with the measurement from SH0ES at $ \leq 1.7 \sigma$. Therefore, the values of $H_0$ for the EDE model are in agreement with SH0ES. We conclude that the EDE model provides a resolution of the Hubble tension.

%%%%%%%%%%%%%%%%%%%%%%%%%%%%%%%%%%%
\vspace*{0.3cm}
\section*{Acknowledgements}
We thank Eiichiro Komatsu for valuable discussions and comments on the draft. We also thank Steen Hannestad, Colin Hill, Emil Holm, Yosuke Kobayashi, Alex Reeves, Th\'eo Simon, Thomas Tram, Sam Witte, and Pierre Zhang for useful discussions and suggestions. LH would like to thank Kavli IPMU for the hospitality where part of this work was conducted. The Kavli IPMU is supported by World Premier International Research Center Initiative (WPI), MEXT, Japan. 

%%%%%%%%%%%%%%%%%%%%%%%%%%%%%%%%%%%

\bibliography{H0_EDE}

%merlin.mbs apsrev4-1.bst 2010-07-25 4.21a (PWD, AO, DPC) hacked
%Control: key (0)
%Control: author (72) initials jnrlst
%Control: editor formatted (1) identically to author
%Control: production of article title (-1) disabled
%Control: page (0) single
%Control: year (1) truncated
%Control: production of eprint (0) enabled
\begin{thebibliography}{56}%
\makeatletter
\providecommand \@ifxundefined [1]{%
 \@ifx{#1\undefined}
}%
\providecommand \@ifnum [1]{%
 \ifnum #1\expandafter \@firstoftwo
 \else \expandafter \@secondoftwo
 \fi
}%
\providecommand \@ifx [1]{%
 \ifx #1\expandafter \@firstoftwo
 \else \expandafter \@secondoftwo
 \fi
}%
\providecommand \natexlab [1]{#1}%
\providecommand \enquote  [1]{``#1''}%
\providecommand \bibnamefont  [1]{#1}%
\providecommand \bibfnamefont [1]{#1}%
\providecommand \citenamefont [1]{#1}%
\providecommand \href@noop [0]{\@secondoftwo}%
\providecommand \href [0]{\begingroup \@sanitize@url \@href}%
\providecommand \@href[1]{\@@startlink{#1}\@@href}%
\providecommand \@@href[1]{\endgroup#1\@@endlink}%
\providecommand \@sanitize@url [0]{\catcode `\\12\catcode `\$12\catcode
  `\&12\catcode `\#12\catcode `\^12\catcode `\_12\catcode `\%12\relax}%
\providecommand \@@startlink[1]{}%
\providecommand \@@endlink[0]{}%
\providecommand \url  [0]{\begingroup\@sanitize@url \@url }%
\providecommand \@url [1]{\endgroup\@href {#1}{\urlprefix }}%
\providecommand \urlprefix  [0]{URL }%
\providecommand \Eprint [0]{\href }%
\providecommand \doibase [0]{http://dx.doi.org/}%
\providecommand \selectlanguage [0]{\@gobble}%
\providecommand \bibinfo  [0]{\@secondoftwo}%
\providecommand \bibfield  [0]{\@secondoftwo}%
\providecommand \translation [1]{[#1]}%
\providecommand \BibitemOpen [0]{}%
\providecommand \bibitemStop [0]{}%
\providecommand \bibitemNoStop [0]{.\EOS\space}%
\providecommand \EOS [0]{\spacefactor3000\relax}%
\providecommand \BibitemShut  [1]{\csname bibitem#1\endcsname}%
\let\auto@bib@innerbib\@empty
%</preamble>
\bibitem [{\citenamefont {Abdalla}\ \emph {et~al.}(2022)\citenamefont {Abdalla}
  \emph {et~al.}}]{Abdalla:2022yfr}%
  \BibitemOpen
  \bibfield  {author} {\bibinfo {author} {\bibfnamefont {E.}~\bibnamefont
  {Abdalla}} \emph {et~al.},\ }\href {\doibase 10.1016/j.jheap.2022.04.002}
  {\bibfield  {journal} {\bibinfo  {journal} {JHEAp}\ }\textbf {\bibinfo
  {volume} {34}},\ \bibinfo {pages} {49} (\bibinfo {year} {2022})},\ \Eprint
  {http://arxiv.org/abs/2203.06142} {arXiv:2203.06142 [astro-ph.CO]}
  \BibitemShut {NoStop}%
\bibitem [{\citenamefont {{Planck Collaboration VI}}(2020)}]{Planck_col_2020}%
  \BibitemOpen
  \bibfield  {author} {\bibinfo {author} {\bibnamefont {{Planck Collaboration
  VI}}},\ }\href {\doibase 10.1051/0004-6361/201833910} {\bibfield  {journal}
  {\bibinfo  {journal} {Astronomy \& Astrophysics}\ }\textbf {\bibinfo {volume}
  {641}},\ \bibinfo {pages} {A6} (\bibinfo {year} {2020})}\BibitemShut
  {NoStop}%
\bibitem [{\citenamefont {Riess}\ \emph {et~al.}(2021)\citenamefont {Riess}
  \emph {et~al.}}]{Riess:2021jrx}%
  \BibitemOpen
  \bibfield  {author} {\bibinfo {author} {\bibfnamefont {A.~G.}\ \bibnamefont
  {Riess}} \emph {et~al.},\ }\href@noop {} {\bibfield  {journal} {\bibinfo
  {journal} {arXiv e-prints}\ } (\bibinfo {year} {2021})},\ \Eprint
  {http://arxiv.org/abs/2112.04510} {arXiv:2112.04510 [astro-ph.CO]}
  \BibitemShut {NoStop}%
\bibitem [{\citenamefont {Poulin}\ \emph {et~al.}(2018)\citenamefont {Poulin},
  \citenamefont {Smith}, \citenamefont {Grin}, \citenamefont {Karwal},\ and\
  \citenamefont {Kamionkowski}}]{Poulin_2018}%
  \BibitemOpen
  \bibfield  {author} {\bibinfo {author} {\bibfnamefont {V.}~\bibnamefont
  {Poulin}}, \bibinfo {author} {\bibfnamefont {T.~L.}\ \bibnamefont {Smith}},
  \bibinfo {author} {\bibfnamefont {D.}~\bibnamefont {Grin}}, \bibinfo {author}
  {\bibfnamefont {T.}~\bibnamefont {Karwal}}, \ and\ \bibinfo {author}
  {\bibfnamefont {M.}~\bibnamefont {Kamionkowski}},\ }\href {\doibase
  10.1103/physrevd.98.083525} {\bibfield  {journal} {\bibinfo  {journal}
  {Physical Review D}\ }\textbf {\bibinfo {volume} {98}} (\bibinfo {year}
  {2018}),\ 10.1103/physrevd.98.083525}\BibitemShut {NoStop}%
\bibitem [{\citenamefont {Poulin}\ \emph {et~al.}(2019)\citenamefont {Poulin},
  \citenamefont {Smith}, \citenamefont {Karwal},\ and\ \citenamefont
  {Kamionkowski}}]{Poulin_2019}%
  \BibitemOpen
  \bibfield  {author} {\bibinfo {author} {\bibfnamefont {V.}~\bibnamefont
  {Poulin}}, \bibinfo {author} {\bibfnamefont {T.~L.}\ \bibnamefont {Smith}},
  \bibinfo {author} {\bibfnamefont {T.}~\bibnamefont {Karwal}}, \ and\ \bibinfo
  {author} {\bibfnamefont {M.}~\bibnamefont {Kamionkowski}},\ }\href {\doibase
  10.1103/physrevlett.122.221301} {\bibfield  {journal} {\bibinfo  {journal}
  {Physical Review Letters}\ }\textbf {\bibinfo {volume} {122}} (\bibinfo
  {year} {2019}),\ 10.1103/physrevlett.122.221301}\BibitemShut {NoStop}%
\bibitem [{\citenamefont {Smith}\ \emph {et~al.}(2020)\citenamefont {Smith},
  \citenamefont {Poulin},\ and\ \citenamefont {Amin}}]{Smith_2020}%
  \BibitemOpen
  \bibfield  {author} {\bibinfo {author} {\bibfnamefont {T.~L.}\ \bibnamefont
  {Smith}}, \bibinfo {author} {\bibfnamefont {V.}~\bibnamefont {Poulin}}, \
  and\ \bibinfo {author} {\bibfnamefont {M.~A.}\ \bibnamefont {Amin}},\ }\href
  {\doibase 10.1103/physrevd.101.063523} {\bibfield  {journal} {\bibinfo
  {journal} {Physical Review D}\ }\textbf {\bibinfo {volume} {101}} (\bibinfo
  {year} {2020}),\ 10.1103/physrevd.101.063523}\BibitemShut {NoStop}%
\bibitem [{\citenamefont {Knox}\ and\ \citenamefont
  {Millea}(2020)}]{Knox:2019rjx}%
  \BibitemOpen
  \bibfield  {author} {\bibinfo {author} {\bibfnamefont {L.}~\bibnamefont
  {Knox}}\ and\ \bibinfo {author} {\bibfnamefont {M.}~\bibnamefont {Millea}},\
  }\href {\doibase 10.1103/PhysRevD.101.043533} {\bibfield  {journal} {\bibinfo
   {journal} {Physical Review D}\ }\textbf {\bibinfo {volume} {101}},\ \bibinfo
  {pages} {043533} (\bibinfo {year} {2020})},\ \Eprint
  {http://arxiv.org/abs/1908.03663} {arXiv:1908.03663 [astro-ph.CO]}
  \BibitemShut {NoStop}%
\bibitem [{\citenamefont {Schöneberg}\ \emph {et~al.}(2021)\citenamefont
  {Schöneberg}, \citenamefont {Abellán}, \citenamefont {Sánchez},
  \citenamefont {Witte}, \citenamefont {Poulin},\ and\ \citenamefont
  {Lesgourgues}}]{Schoeneberg_2021_H0}%
  \BibitemOpen
  \bibfield  {author} {\bibinfo {author} {\bibfnamefont {N.}~\bibnamefont
  {Schöneberg}}, \bibinfo {author} {\bibfnamefont {G.~F.}\ \bibnamefont
  {Abellán}}, \bibinfo {author} {\bibfnamefont {A.~P.}\ \bibnamefont
  {Sánchez}}, \bibinfo {author} {\bibfnamefont {S.~J.}\ \bibnamefont {Witte}},
  \bibinfo {author} {\bibfnamefont {V.}~\bibnamefont {Poulin}}, \ and\ \bibinfo
  {author} {\bibfnamefont {J.}~\bibnamefont {Lesgourgues}},\ }\href@noop {}
  {\bibfield  {journal} {\bibinfo  {journal} {arXiv e-prints}\ } (\bibinfo
  {year} {2021})},\ \Eprint {http://arxiv.org/abs/2107.10291} {arXiv:2107.10291
  [astro-ph.CO]} \BibitemShut {NoStop}%
\bibitem [{\citenamefont {Hill}\ \emph {et~al.}(2020)\citenamefont {Hill},
  \citenamefont {McDonough}, \citenamefont {Toomey},\ and\ \citenamefont
  {Alexander}}]{Hill_2020}%
  \BibitemOpen
  \bibfield  {author} {\bibinfo {author} {\bibfnamefont {J.~C.}\ \bibnamefont
  {Hill}}, \bibinfo {author} {\bibfnamefont {E.}~\bibnamefont {McDonough}},
  \bibinfo {author} {\bibfnamefont {M.~W.}\ \bibnamefont {Toomey}}, \ and\
  \bibinfo {author} {\bibfnamefont {S.}~\bibnamefont {Alexander}},\ }\href
  {\doibase 10.1103/physrevd.102.043507} {\bibfield  {journal} {\bibinfo
  {journal} {Physical Review D}\ }\textbf {\bibinfo {volume} {102}} (\bibinfo
  {year} {2020}),\ 10.1103/physrevd.102.043507}\BibitemShut {NoStop}%
\bibitem [{\citenamefont {Ivanov}\ \emph
  {et~al.}(2020{\natexlab{a}})\citenamefont {Ivanov}, \citenamefont
  {McDonough}, \citenamefont {Hill}, \citenamefont {Simonović}, \citenamefont
  {Toomey}, \citenamefont {Alexander},\ and\ \citenamefont
  {Zaldarriaga}}]{Ivanov_2020}%
  \BibitemOpen
  \bibfield  {author} {\bibinfo {author} {\bibfnamefont {M.~M.}\ \bibnamefont
  {Ivanov}}, \bibinfo {author} {\bibfnamefont {E.}~\bibnamefont {McDonough}},
  \bibinfo {author} {\bibfnamefont {J.~C.}\ \bibnamefont {Hill}}, \bibinfo
  {author} {\bibfnamefont {M.}~\bibnamefont {Simonović}}, \bibinfo {author}
  {\bibfnamefont {M.~W.}\ \bibnamefont {Toomey}}, \bibinfo {author}
  {\bibfnamefont {S.}~\bibnamefont {Alexander}}, \ and\ \bibinfo {author}
  {\bibfnamefont {M.}~\bibnamefont {Zaldarriaga}},\ }\href {\doibase
  10.1103/physrevd.102.103502} {\bibfield  {journal} {\bibinfo  {journal}
  {Physical Review D}\ }\textbf {\bibinfo {volume} {102}} (\bibinfo {year}
  {2020}{\natexlab{a}}),\ 10.1103/physrevd.102.103502}\BibitemShut {NoStop}%
\bibitem [{\citenamefont {D’Amico}\ \emph {et~al.}(2021)\citenamefont
  {D’Amico}, \citenamefont {Senatore}, \citenamefont {Zhang},\ and\
  \citenamefont {Zheng}}]{DAmico_2021}%
  \BibitemOpen
  \bibfield  {author} {\bibinfo {author} {\bibfnamefont {G.}~\bibnamefont
  {D’Amico}}, \bibinfo {author} {\bibfnamefont {L.}~\bibnamefont {Senatore}},
  \bibinfo {author} {\bibfnamefont {P.}~\bibnamefont {Zhang}}, \ and\ \bibinfo
  {author} {\bibfnamefont {H.}~\bibnamefont {Zheng}},\ }\href {\doibase
  10.1088/1475-7516/2021/05/072} {\bibfield  {journal} {\bibinfo  {journal}
  {Journal of Cosmology and Astroparticle Physics}\ }\textbf {\bibinfo {volume}
  {2021}},\ \bibinfo {pages} {072} (\bibinfo {year} {2021})}\BibitemShut
  {NoStop}%
\bibitem [{\citenamefont {Secco}\ \emph {et~al.}(2022)\citenamefont {Secco},
  \citenamefont {Karwal}, \citenamefont {Hu},\ and\ \citenamefont
  {Krause}}]{Secco:2022kqg}%
  \BibitemOpen
  \bibfield  {author} {\bibinfo {author} {\bibfnamefont {L.~F.}\ \bibnamefont
  {Secco}}, \bibinfo {author} {\bibfnamefont {T.}~\bibnamefont {Karwal}},
  \bibinfo {author} {\bibfnamefont {W.}~\bibnamefont {Hu}}, \ and\ \bibinfo
  {author} {\bibfnamefont {E.}~\bibnamefont {Krause}},\ }\href@noop {}
  {\bibfield  {journal} {\bibinfo  {journal} {arXiv e-prints}\ } (\bibinfo
  {year} {2022})},\ \Eprint {http://arxiv.org/abs/2209.12997} {arXiv:2209.12997
  [astro-ph.CO]} \BibitemShut {NoStop}%
\bibitem [{\citenamefont {Herold}\ \emph {et~al.}(2022)\citenamefont {Herold},
  \citenamefont {Ferreira},\ and\ \citenamefont {Komatsu}}]{Herold:2021ksg}%
  \BibitemOpen
  \bibfield  {author} {\bibinfo {author} {\bibfnamefont {L.}~\bibnamefont
  {Herold}}, \bibinfo {author} {\bibfnamefont {E.~G.~M.}\ \bibnamefont
  {Ferreira}}, \ and\ \bibinfo {author} {\bibfnamefont {E.}~\bibnamefont
  {Komatsu}},\ }\href {\doibase 10.3847/2041-8213/ac63a3} {\bibfield  {journal}
  {\bibinfo  {journal} {Astrophys. J. Lett.}\ }\textbf {\bibinfo {volume}
  {929}},\ \bibinfo {pages} {L16} (\bibinfo {year} {2022})},\ \Eprint
  {http://arxiv.org/abs/2112.12140} {arXiv:2112.12140 [astro-ph.CO]}
  \BibitemShut {NoStop}%
\bibitem [{\citenamefont {Niedermann}\ and\ \citenamefont
  {Sloth}(2020)}]{Niedermann_2020}%
  \BibitemOpen
  \bibfield  {author} {\bibinfo {author} {\bibfnamefont {F.}~\bibnamefont
  {Niedermann}}\ and\ \bibinfo {author} {\bibfnamefont {M.~S.}\ \bibnamefont
  {Sloth}},\ }\href {\doibase 10.1103/physrevd.102.063527} {\bibfield
  {journal} {\bibinfo  {journal} {Physical Review D}\ }\textbf {\bibinfo
  {volume} {102}} (\bibinfo {year} {2020}),\
  10.1103/physrevd.102.063527}\BibitemShut {NoStop}%
\bibitem [{\citenamefont {Murgia}\ \emph {et~al.}(2021)\citenamefont {Murgia},
  \citenamefont {Abell\'an},\ and\ \citenamefont {Poulin}}]{Murgia:2020ryi}%
  \BibitemOpen
  \bibfield  {author} {\bibinfo {author} {\bibfnamefont {R.}~\bibnamefont
  {Murgia}}, \bibinfo {author} {\bibfnamefont {G.~F.}\ \bibnamefont
  {Abell\'an}}, \ and\ \bibinfo {author} {\bibfnamefont {V.}~\bibnamefont
  {Poulin}},\ }\href {\doibase 10.1103/PhysRevD.103.063502} {\bibfield
  {journal} {\bibinfo  {journal} {Phys. Rev. D}\ }\textbf {\bibinfo {volume}
  {103}},\ \bibinfo {pages} {063502} (\bibinfo {year} {2021})},\ \Eprint
  {http://arxiv.org/abs/2009.10733} {arXiv:2009.10733 [astro-ph.CO]}
  \BibitemShut {NoStop}%
\bibitem [{\citenamefont {Smith}\ \emph {et~al.}(2021)\citenamefont {Smith},
  \citenamefont {Poulin}, \citenamefont {Bernal}, \citenamefont {Boddy},
  \citenamefont {Kamionkowski},\ and\ \citenamefont {Murgia}}]{Smith:2020rxx}%
  \BibitemOpen
  \bibfield  {author} {\bibinfo {author} {\bibfnamefont {T.~L.}\ \bibnamefont
  {Smith}}, \bibinfo {author} {\bibfnamefont {V.}~\bibnamefont {Poulin}},
  \bibinfo {author} {\bibfnamefont {J.~L.}\ \bibnamefont {Bernal}}, \bibinfo
  {author} {\bibfnamefont {K.~K.}\ \bibnamefont {Boddy}}, \bibinfo {author}
  {\bibfnamefont {M.}~\bibnamefont {Kamionkowski}}, \ and\ \bibinfo {author}
  {\bibfnamefont {R.}~\bibnamefont {Murgia}},\ }\href {\doibase
  10.1103/PhysRevD.103.123542} {\bibfield  {journal} {\bibinfo  {journal}
  {Phys. Rev. D}\ }\textbf {\bibinfo {volume} {103}},\ \bibinfo {pages}
  {123542} (\bibinfo {year} {2021})},\ \Eprint
  {http://arxiv.org/abs/2009.10740} {arXiv:2009.10740 [astro-ph.CO]}
  \BibitemShut {NoStop}%
\bibitem [{\citenamefont {{G{\'o}mez-Valent}}(2022)}]{Gomez-Valent:2022hkb}%
  \BibitemOpen
  \bibfield  {author} {\bibinfo {author} {\bibfnamefont {A.}~\bibnamefont
  {{G{\'o}mez-Valent}}},\ }\href@noop {} {\bibfield  {journal} {\bibinfo
  {journal} {arXiv e-prints}\ ,\ \bibinfo {eid} {arXiv:2203.16285}} (\bibinfo
  {year} {2022})},\ \Eprint {http://arxiv.org/abs/2203.16285} {arXiv:2203.16285
  [astro-ph.CO]} \BibitemShut {NoStop}%
\bibitem [{\citenamefont {Hogg}\ \emph {et~al.}(2013)\citenamefont {Hogg},
  \citenamefont {McKean},\ and\ \citenamefont {Craig}}]{hogg2013introduction}%
  \BibitemOpen
  \bibfield  {author} {\bibinfo {author} {\bibfnamefont {R.}~\bibnamefont
  {Hogg}}, \bibinfo {author} {\bibfnamefont {J.}~\bibnamefont {McKean}}, \ and\
  \bibinfo {author} {\bibfnamefont {A.}~\bibnamefont {Craig}},\ }\href
  {https://books.google.co.jp/books?id=xFlVYAAACAAJ} {\emph {\bibinfo {title}
  {Introduction to Mathematical Statistics}}},\ \bibinfo {edition} {7th}\ ed.,\
  Always learning\ (\bibinfo  {publisher} {Pearson},\ \bibinfo {year} {2013})\
  \bibinfo {note} {. Theorem 6.1.2}\BibitemShut {NoStop}%
\bibitem [{\citenamefont {Ivanov}\ \emph
  {et~al.}(2020{\natexlab{b}})\citenamefont {Ivanov}, \citenamefont
  {Simonović},\ and\ \citenamefont {Zaldarriaga}}]{Ivanov_2020_full-shape}%
  \BibitemOpen
  \bibfield  {author} {\bibinfo {author} {\bibfnamefont {M.~M.}\ \bibnamefont
  {Ivanov}}, \bibinfo {author} {\bibfnamefont {M.}~\bibnamefont {Simonović}},
  \ and\ \bibinfo {author} {\bibfnamefont {M.}~\bibnamefont {Zaldarriaga}},\
  }\href {\doibase 10.1088/1475-7516/2020/05/042} {\bibfield  {journal}
  {\bibinfo  {journal} {Journal of Cosmology and Astroparticle Physics}\
  }\textbf {\bibinfo {volume} {2020}},\ \bibinfo {pages} {042–042} (\bibinfo
  {year} {2020}{\natexlab{b}})}\BibitemShut {NoStop}%
\bibitem [{\citenamefont {d’ Amico}\ \emph {et~al.}(2020)\citenamefont {d’
  Amico}, \citenamefont {Gleyzes}, \citenamefont {Kokron}, \citenamefont
  {Markovic}, \citenamefont {Senatore}, \citenamefont {Zhang}, \citenamefont
  {Beutler},\ and\ \citenamefont {Gil-Marín}}]{DAmico_2020}%
  \BibitemOpen
  \bibfield  {author} {\bibinfo {author} {\bibfnamefont {G.}~\bibnamefont {d’
  Amico}}, \bibinfo {author} {\bibfnamefont {J.}~\bibnamefont {Gleyzes}},
  \bibinfo {author} {\bibfnamefont {N.}~\bibnamefont {Kokron}}, \bibinfo
  {author} {\bibfnamefont {K.}~\bibnamefont {Markovic}}, \bibinfo {author}
  {\bibfnamefont {L.}~\bibnamefont {Senatore}}, \bibinfo {author}
  {\bibfnamefont {P.}~\bibnamefont {Zhang}}, \bibinfo {author} {\bibfnamefont
  {F.}~\bibnamefont {Beutler}}, \ and\ \bibinfo {author} {\bibfnamefont
  {H.}~\bibnamefont {Gil-Marín}},\ }\href {\doibase
  10.1088/1475-7516/2020/05/005} {\bibfield  {journal} {\bibinfo  {journal}
  {Journal of Cosmology and Astroparticle Physics}\ }\textbf {\bibinfo {volume}
  {2020}},\ \bibinfo {pages} {005–005} (\bibinfo {year} {2020})}\BibitemShut
  {NoStop}%
\bibitem [{\citenamefont {Reeves}\ \emph {et~al.}(2022)\citenamefont {Reeves},
  \citenamefont {Herold}, \citenamefont {Vagnozzi}, \citenamefont {Sherwin},\
  and\ \citenamefont {Ferreira}}]{Reeves:2022aoi}%
  \BibitemOpen
  \bibfield  {author} {\bibinfo {author} {\bibfnamefont {A.}~\bibnamefont
  {Reeves}}, \bibinfo {author} {\bibfnamefont {L.}~\bibnamefont {Herold}},
  \bibinfo {author} {\bibfnamefont {S.}~\bibnamefont {Vagnozzi}}, \bibinfo
  {author} {\bibfnamefont {B.~D.}\ \bibnamefont {Sherwin}}, \ and\ \bibinfo
  {author} {\bibfnamefont {E.~G.~M.}\ \bibnamefont {Ferreira}},\ }\href@noop {}
  {\bibfield  {journal} {\bibinfo  {journal} {arXiv e-prints}\ } (\bibinfo
  {year} {2022})},\ \Eprint {http://arxiv.org/abs/2207.01501} {arXiv:2207.01501
  [astro-ph.CO]} \BibitemShut {NoStop}%
\bibitem [{\citenamefont {Hamann}(2012)}]{Hamann_2012}%
  \BibitemOpen
  \bibfield  {author} {\bibinfo {author} {\bibfnamefont {J.}~\bibnamefont
  {Hamann}},\ }\href {\doibase 10.1088/1475-7516/2012/03/021} {\bibfield
  {journal} {\bibinfo  {journal} {Journal of Cosmology and Astroparticle
  Physics}\ }\textbf {\bibinfo {volume} {2012}},\ \bibinfo {pages} {021}
  (\bibinfo {year} {2012})}\BibitemShut {NoStop}%
\bibitem [{\citenamefont {{Planck Collaboration Int.
  XVI}}(2014)}]{Planck_col_2014}%
  \BibitemOpen
  \bibfield  {author} {\bibinfo {author} {\bibnamefont {{Planck Collaboration
  Int. XVI}}},\ }\href {\doibase 10.1051/0004-6361/201323003} {\bibfield
  {journal} {\bibinfo  {journal} {Astronomy \& Astrophysics}\ }\textbf
  {\bibinfo {volume} {566}},\ \bibinfo {pages} {A54} (\bibinfo {year}
  {2014})}\BibitemShut {NoStop}%
\bibitem [{\citenamefont {Campeti}\ \emph {et~al.}(2022)\citenamefont
  {Campeti}, \citenamefont {Özsoy}, \citenamefont {Obata},\ and\ \citenamefont
  {Shiraishi}}]{Campeti_2022}%
  \BibitemOpen
  \bibfield  {author} {\bibinfo {author} {\bibfnamefont {P.}~\bibnamefont
  {Campeti}}, \bibinfo {author} {\bibfnamefont {O.}~\bibnamefont {Özsoy}},
  \bibinfo {author} {\bibfnamefont {I.}~\bibnamefont {Obata}}, \ and\ \bibinfo
  {author} {\bibfnamefont {M.}~\bibnamefont {Shiraishi}},\ }\href {\doibase
  10.1088/1475-7516/2022/07/039} {\bibfield  {journal} {\bibinfo  {journal}
  {Journal of Cosmology and Astroparticle Physics}\ }\textbf {\bibinfo {volume}
  {2022}},\ \bibinfo {pages} {039} (\bibinfo {year} {2022})}\BibitemShut
  {NoStop}%
\bibitem [{\citenamefont {Campeti}\ and\ \citenamefont
  {Komatsu}(2022)}]{CampetiKomatsu_2022}%
  \BibitemOpen
  \bibfield  {author} {\bibinfo {author} {\bibfnamefont {P.}~\bibnamefont
  {Campeti}}\ and\ \bibinfo {author} {\bibfnamefont {E.}~\bibnamefont
  {Komatsu}},\ }\href {\doibase 10.48550/ARXIV.2205.05617} {\bibfield
  {journal} {\bibinfo  {journal} {arXiv e-prints}\ } (\bibinfo {year} {2022}),\
  10.48550/ARXIV.2205.05617}\BibitemShut {NoStop}%
\bibitem [{\citenamefont {Kamionkowski}\ \emph {et~al.}(2014)\citenamefont
  {Kamionkowski}, \citenamefont {Pradler},\ and\ \citenamefont
  {Walker}}]{kamionkowski_2014}%
  \BibitemOpen
  \bibfield  {author} {\bibinfo {author} {\bibfnamefont {M.}~\bibnamefont
  {Kamionkowski}}, \bibinfo {author} {\bibfnamefont {J.}~\bibnamefont
  {Pradler}}, \ and\ \bibinfo {author} {\bibfnamefont {D.~G.~E.}\ \bibnamefont
  {Walker}},\ }\href {\doibase 10.1103/PhysRevLett.113.251302} {\bibfield
  {journal} {\bibinfo  {journal} {Physical Review Letters}\ }\textbf {\bibinfo
  {volume} {113}},\ \bibinfo {pages} {251302} (\bibinfo {year} {2014})},\
  \Eprint {http://arxiv.org/abs/1409.0549} {arXiv:1409.0549 [hep-ph]}
  \BibitemShut {NoStop}%
\bibitem [{\citenamefont {Karwal}\ and\ \citenamefont
  {Kamionkowski}(2016)}]{Karwal:2016vyq}%
  \BibitemOpen
  \bibfield  {author} {\bibinfo {author} {\bibfnamefont {T.}~\bibnamefont
  {Karwal}}\ and\ \bibinfo {author} {\bibfnamefont {M.}~\bibnamefont
  {Kamionkowski}},\ }\href {\doibase 10.1103/PhysRevD.94.103523} {\bibfield
  {journal} {\bibinfo  {journal} {Physical Review D}\ }\textbf {\bibinfo
  {volume} {94}},\ \bibinfo {pages} {103523} (\bibinfo {year} {2016})},\
  \Eprint {http://arxiv.org/abs/1608.01309} {arXiv:1608.01309 [astro-ph.CO]}
  \BibitemShut {NoStop}%
\bibitem [{\citenamefont {Caldwell}\ and\ \citenamefont
  {Devulder}(2018)}]{Caldwell_2018}%
  \BibitemOpen
  \bibfield  {author} {\bibinfo {author} {\bibfnamefont {R.~R.}\ \bibnamefont
  {Caldwell}}\ and\ \bibinfo {author} {\bibfnamefont {C.}~\bibnamefont
  {Devulder}},\ }\href {\doibase 10.1103/PhysRevD.97.023532} {\bibfield
  {journal} {\bibinfo  {journal} {Physical Review D}\ }\textbf {\bibinfo
  {volume} {97}},\ \bibinfo {pages} {023532} (\bibinfo {year} {2018})},\
  \Eprint {http://arxiv.org/abs/1706.03765} {arXiv:1706.03765 [astro-ph.CO]}
  \BibitemShut {NoStop}%
\bibitem [{\citenamefont {Bernal}\ \emph {et~al.}(2016)\citenamefont {Bernal},
  \citenamefont {Verde},\ and\ \citenamefont {Riess}}]{Bernal_2016}%
  \BibitemOpen
  \bibfield  {author} {\bibinfo {author} {\bibfnamefont {J.~L.}\ \bibnamefont
  {Bernal}}, \bibinfo {author} {\bibfnamefont {L.}~\bibnamefont {Verde}}, \
  and\ \bibinfo {author} {\bibfnamefont {A.~G.}\ \bibnamefont {Riess}},\ }\href
  {\doibase 10.1088/1475-7516/2016/10/019} {\bibfield  {journal} {\bibinfo
  {journal} {Journal of Cosmology and Astroparticle Physics}\ }\textbf
  {\bibinfo {volume} {10}},\ \bibinfo {pages} {019} (\bibinfo {year} {2016})},\
  \Eprint {http://arxiv.org/abs/1607.05617} {arXiv:1607.05617 [astro-ph.CO]}
  \BibitemShut {NoStop}%
\bibitem [{\citenamefont
  {\url{https://github.com/Michalychforever/EDE_class_pt}}()}]{class_EDE_PT}%
  \BibitemOpen
  \bibfield  {author} {\bibinfo {author} {\bibnamefont
  {\url{https://github.com/Michalychforever/EDE_class_pt}}},\ }\href@noop {}
  {}\BibitemShut {NoStop}%
\bibitem [{\citenamefont {Lesgourgues}(2011)}]{Lesgourgues_2011}%
  \BibitemOpen
  \bibfield  {author} {\bibinfo {author} {\bibfnamefont {J.}~\bibnamefont
  {Lesgourgues}},\ }\href@noop {} {\bibfield  {journal} {\bibinfo  {journal}
  {arXiv e-prints}\ } (\bibinfo {year} {2011})},\ \Eprint
  {http://arxiv.org/abs/1104.2932} {arXiv:1104.2932 [astro-ph.IM]} \BibitemShut
  {NoStop}%
\bibitem [{\citenamefont {Blas}\ \emph {et~al.}(2011)\citenamefont {Blas},
  \citenamefont {Lesgourgues},\ and\ \citenamefont {Tram}}]{Blas_2011}%
  \BibitemOpen
  \bibfield  {author} {\bibinfo {author} {\bibfnamefont {D.}~\bibnamefont
  {Blas}}, \bibinfo {author} {\bibfnamefont {J.}~\bibnamefont {Lesgourgues}}, \
  and\ \bibinfo {author} {\bibfnamefont {T.}~\bibnamefont {Tram}},\ }\href
  {\doibase 10.1088/1475-7516/2011/07/034} {\bibfield  {journal} {\bibinfo
  {journal} {Journal of Cosmology and Astroparticle Physics}\ }\textbf
  {\bibinfo {volume} {2011}},\ \bibinfo {pages} {034–034} (\bibinfo {year}
  {2011})}\BibitemShut {NoStop}%
\bibitem [{\citenamefont {Chudaykin}\ \emph {et~al.}(2020)\citenamefont
  {Chudaykin}, \citenamefont {Ivanov}, \citenamefont {Philcox},\ and\
  \citenamefont {Simonović}}]{Chudaykin_2020}%
  \BibitemOpen
  \bibfield  {author} {\bibinfo {author} {\bibfnamefont {A.}~\bibnamefont
  {Chudaykin}}, \bibinfo {author} {\bibfnamefont {M.~M.}\ \bibnamefont
  {Ivanov}}, \bibinfo {author} {\bibfnamefont {O.~H.}\ \bibnamefont {Philcox}},
  \ and\ \bibinfo {author} {\bibfnamefont {M.}~\bibnamefont {Simonović}},\
  }\href {\doibase 10.1103/physrevd.102.063533} {\bibfield  {journal} {\bibinfo
   {journal} {Physical Review D}\ }\textbf {\bibinfo {volume} {102}} (\bibinfo
  {year} {2020}),\ 10.1103/physrevd.102.063533}\BibitemShut {NoStop}%
\bibitem [{\citenamefont {Alam}\ \emph {et~al.}(2017)\citenamefont {Alam},
  \citenamefont {Ata}, \citenamefont {Bailey}, \citenamefont {Beutler},
  \citenamefont {Bizyaev}, \citenamefont {Blazek}, \citenamefont {Bolton},
  \citenamefont {Brownstein}, \citenamefont {Burden}, \citenamefont {Chuang},\
  and\ \citenamefont {et~al.}}]{BOSS_col_2017}%
  \BibitemOpen
  \bibfield  {author} {\bibinfo {author} {\bibfnamefont {S.}~\bibnamefont
  {Alam}}, \bibinfo {author} {\bibfnamefont {M.}~\bibnamefont {Ata}}, \bibinfo
  {author} {\bibfnamefont {S.}~\bibnamefont {Bailey}}, \bibinfo {author}
  {\bibfnamefont {F.}~\bibnamefont {Beutler}}, \bibinfo {author} {\bibfnamefont
  {D.}~\bibnamefont {Bizyaev}}, \bibinfo {author} {\bibfnamefont {J.~A.}\
  \bibnamefont {Blazek}}, \bibinfo {author} {\bibfnamefont {A.~S.}\
  \bibnamefont {Bolton}}, \bibinfo {author} {\bibfnamefont {J.~R.}\
  \bibnamefont {Brownstein}}, \bibinfo {author} {\bibfnamefont
  {A.}~\bibnamefont {Burden}}, \bibinfo {author} {\bibfnamefont {C.-H.}\
  \bibnamefont {Chuang}}, \ and\ \bibinfo {author} {\bibnamefont {et~al.}},\
  }\href {\doibase 10.1093/mnras/stx721} {\bibfield  {journal} {\bibinfo
  {journal} {Monthly Notices of the Royal Astronomical Society}\ }\textbf
  {\bibinfo {volume} {470}},\ \bibinfo {pages} {2617–2652} (\bibinfo {year}
  {2017})}\BibitemShut {NoStop}%
\bibitem [{\citenamefont {Baumann}\ \emph {et~al.}(2012)\citenamefont
  {Baumann}, \citenamefont {Nicolis}, \citenamefont {Senatore},\ and\
  \citenamefont {Zaldarriaga}}]{Baumann_2012}%
  \BibitemOpen
  \bibfield  {author} {\bibinfo {author} {\bibfnamefont {D.}~\bibnamefont
  {Baumann}}, \bibinfo {author} {\bibfnamefont {A.}~\bibnamefont {Nicolis}},
  \bibinfo {author} {\bibfnamefont {L.}~\bibnamefont {Senatore}}, \ and\
  \bibinfo {author} {\bibfnamefont {M.}~\bibnamefont {Zaldarriaga}},\ }\href
  {\doibase 10.1088/1475-7516/2012/07/051} {\bibfield  {journal} {\bibinfo
  {journal} {Journal of Cosmology and Astroparticle Physics}\ }\textbf
  {\bibinfo {volume} {2012}},\ \bibinfo {pages} {051–051} (\bibinfo {year}
  {2012})}\BibitemShut {NoStop}%
\bibitem [{\citenamefont {Carrasco}\ \emph {et~al.}(2012)\citenamefont
  {Carrasco}, \citenamefont {Hertzberg},\ and\ \citenamefont
  {Senatore}}]{Carrasco_2012}%
  \BibitemOpen
  \bibfield  {author} {\bibinfo {author} {\bibfnamefont {J.~J.~M.}\
  \bibnamefont {Carrasco}}, \bibinfo {author} {\bibfnamefont {M.~P.}\
  \bibnamefont {Hertzberg}}, \ and\ \bibinfo {author} {\bibfnamefont
  {L.}~\bibnamefont {Senatore}},\ }\href {\doibase 10.1007/jhep09(2012)082}
  {\bibfield  {journal} {\bibinfo  {journal} {Journal of High Energy Physics}\
  }\textbf {\bibinfo {volume} {2012}} (\bibinfo {year} {2012}),\
  10.1007/jhep09(2012)082}\BibitemShut {NoStop}%
\bibitem [{\citenamefont {Beutler}\ and\ \citenamefont
  {McDonald}(2021)}]{Beutler:2021eqq}%
  \BibitemOpen
  \bibfield  {author} {\bibinfo {author} {\bibfnamefont {F.}~\bibnamefont
  {Beutler}}\ and\ \bibinfo {author} {\bibfnamefont {P.}~\bibnamefont
  {McDonald}},\ }\href {\doibase 10.1088/1475-7516/2021/11/031} {\bibfield
  {journal} {\bibinfo  {journal} {Journal of Cosmology and Astroparticle
  Physics}\ }\textbf {\bibinfo {volume} {11}},\ \bibinfo {pages} {031}
  (\bibinfo {year} {2021})},\ \Eprint {http://arxiv.org/abs/2106.06324}
  {arXiv:2106.06324 [astro-ph.CO]} \BibitemShut {NoStop}%
\bibitem [{\citenamefont {Abbott}\ \emph {et~al.}(2022)\citenamefont {Abbott},
  \citenamefont {Aguena}, \citenamefont {Alarcon}, \citenamefont {Allam},\ and\
  \citenamefont {Alves}}]{Abbott_2022}%
  \BibitemOpen
  \bibfield  {author} {\bibinfo {author} {\bibfnamefont {T.}~\bibnamefont
  {Abbott}}, \bibinfo {author} {\bibfnamefont {M.}~\bibnamefont {Aguena}},
  \bibinfo {author} {\bibfnamefont {A.}~\bibnamefont {Alarcon}}, \bibinfo
  {author} {\bibfnamefont {S.}~\bibnamefont {Allam}}, \ and\ \bibinfo {author}
  {\bibfnamefont {O.}~\bibnamefont {Alves}},\ }\href {\doibase
  10.1103/physrevd.105.023520} {\bibfield  {journal} {\bibinfo  {journal}
  {Physical Review D}\ }\textbf {\bibinfo {volume} {105}} (\bibinfo {year}
  {2022}),\ 10.1103/physrevd.105.023520}\BibitemShut {NoStop}%
\bibitem [{\citenamefont {Brinckmann}\ and\ \citenamefont
  {Lesgourgues}(2018)}]{Brinckmann_2018}%
  \BibitemOpen
  \bibfield  {author} {\bibinfo {author} {\bibfnamefont {T.}~\bibnamefont
  {Brinckmann}}\ and\ \bibinfo {author} {\bibfnamefont {J.}~\bibnamefont
  {Lesgourgues}},\ }\href@noop {} {\bibfield  {journal} {\bibinfo  {journal}
  {arXiv e-prints}\ } (\bibinfo {year} {2018})},\ \Eprint
  {http://arxiv.org/abs/1804.07261} {arXiv:1804.07261 [astro-ph.CO]}
  \BibitemShut {NoStop}%
\bibitem [{\citenamefont {{Metropolis}}\ \emph {et~al.}(1953)\citenamefont
  {{Metropolis}}, \citenamefont {{Rosenbluth}}, \citenamefont {{Rosenbluth}},
  \citenamefont {{Teller}},\ and\ \citenamefont {{Teller}}}]{Metropolis_1953}%
  \BibitemOpen
  \bibfield  {author} {\bibinfo {author} {\bibfnamefont {N.}~\bibnamefont
  {{Metropolis}}}, \bibinfo {author} {\bibfnamefont {A.~W.}\ \bibnamefont
  {{Rosenbluth}}}, \bibinfo {author} {\bibfnamefont {M.~N.}\ \bibnamefont
  {{Rosenbluth}}}, \bibinfo {author} {\bibfnamefont {A.~H.}\ \bibnamefont
  {{Teller}}}, \ and\ \bibinfo {author} {\bibfnamefont {E.}~\bibnamefont
  {{Teller}}},\ }\href {\doibase 10.1063/1.1699114} {\bibfield  {journal}
  {\bibinfo  {journal} {The Journal of Chemical Physics}\ }\textbf {\bibinfo
  {volume} {21}},\ \bibinfo {pages} {1087} (\bibinfo {year}
  {1953})}\BibitemShut {NoStop}%
\bibitem [{\citenamefont {{Hastings}}(1970)}]{Hastings_1970}%
  \BibitemOpen
  \bibfield  {author} {\bibinfo {author} {\bibfnamefont {W.~K.}\ \bibnamefont
  {{Hastings}}},\ }\href {\doibase 10.1093/biomet/57.1.97} {\bibfield
  {journal} {\bibinfo  {journal} {Biometrika}\ }\textbf {\bibinfo {volume}
  {57}},\ \bibinfo {pages} {97} (\bibinfo {year} {1970})}\BibitemShut {NoStop}%
\bibitem [{\citenamefont {Philcox}\ and\ \citenamefont
  {Ivanov}(2022)}]{Philcox_2022}%
  \BibitemOpen
  \bibfield  {author} {\bibinfo {author} {\bibfnamefont {O.~H.~E.}\
  \bibnamefont {Philcox}}\ and\ \bibinfo {author} {\bibfnamefont {M.~M.}\
  \bibnamefont {Ivanov}},\ }\href {\doibase 10.1103/physrevd.105.043517}
  {\bibfield  {journal} {\bibinfo  {journal} {Physical Review D}\ }\textbf
  {\bibinfo {volume} {105}} (\bibinfo {year} {2022}),\
  10.1103/physrevd.105.043517}\BibitemShut {NoStop}%
\bibitem [{\citenamefont {Hannestad}(1999)}]{Hannestad_1999}%
  \BibitemOpen
  \bibfield  {author} {\bibinfo {author} {\bibfnamefont {S.}~\bibnamefont
  {Hannestad}},\ }\href {\doibase 10.1103/physrevd.61.023002} {\bibfield
  {journal} {\bibinfo  {journal} {Physical Review D}\ }\textbf {\bibinfo
  {volume} {61}} (\bibinfo {year} {1999}),\
  10.1103/physrevd.61.023002}\BibitemShut {NoStop}%
\bibitem [{\citenamefont {Feldman}\ and\ \citenamefont
  {Cousins}(1998)}]{Feldman_1998}%
  \BibitemOpen
  \bibfield  {author} {\bibinfo {author} {\bibfnamefont {G.~J.}\ \bibnamefont
  {Feldman}}\ and\ \bibinfo {author} {\bibfnamefont {R.~D.}\ \bibnamefont
  {Cousins}},\ }\href {\doibase 10.1103/physrevd.57.3873} {\bibfield  {journal}
  {\bibinfo  {journal} {Physical Review D}\ }\textbf {\bibinfo {volume} {57}},\
  \bibinfo {pages} {3873–3889} (\bibinfo {year} {1998})}\BibitemShut
  {NoStop}%
\bibitem [{\citenamefont {Neyman}(1937)}]{Neyman:1937uhy}%
  \BibitemOpen
  \bibfield  {author} {\bibinfo {author} {\bibfnamefont {J.}~\bibnamefont
  {Neyman}},\ }\href {\doibase 10.1098/rsta.1937.0005} {\bibfield  {journal}
  {\bibinfo  {journal} {Philosophical Transactions of the Royal Society of
  London A}\ }\textbf {\bibinfo {volume} {236}},\ \bibinfo {pages} {333}
  (\bibinfo {year} {1937})}\BibitemShut {NoStop}%
\bibitem [{\citenamefont {Burnham}\ and\ \citenamefont
  {Anderson}(2002)}]{burnham2002model}%
  \BibitemOpen
  \bibfield  {author} {\bibinfo {author} {\bibfnamefont {K.}~\bibnamefont
  {Burnham}}\ and\ \bibinfo {author} {\bibfnamefont {D.}~\bibnamefont
  {Anderson}},\ }\href@noop {} {\emph {\bibinfo {title} {Model selection and
  multimodel inference: a practical information-theoretic approach}}}\
  (\bibinfo  {publisher} {Springer Verlag},\ \bibinfo {year}
  {2002})\BibitemShut {NoStop}%
\bibitem [{\citenamefont {Simon}\ \emph {et~al.}(2022)\citenamefont {Simon},
  \citenamefont {Zhang}, \citenamefont {Poulin},\ and\ \citenamefont
  {Smith}}]{Simon_2022ede}%
  \BibitemOpen
  \bibfield  {author} {\bibinfo {author} {\bibfnamefont {T.}~\bibnamefont
  {Simon}}, \bibinfo {author} {\bibfnamefont {P.}~\bibnamefont {Zhang}},
  \bibinfo {author} {\bibfnamefont {V.}~\bibnamefont {Poulin}}, \ and\ \bibinfo
  {author} {\bibfnamefont {T.~L.}\ \bibnamefont {Smith}},\ }\href {\doibase
  10.48550/ARXIV.2208.05930} {\bibfield  {journal} {\bibinfo  {journal} {arXiv
  e-prints}\ } (\bibinfo {year} {2022}),\
  10.48550/ARXIV.2208.05930}\BibitemShut {NoStop}%
\bibitem [{\citenamefont {Vagnozzi}(2021)}]{Vagnozzi_2021}%
  \BibitemOpen
  \bibfield  {author} {\bibinfo {author} {\bibfnamefont {S.}~\bibnamefont
  {Vagnozzi}},\ }\href {\doibase 10.1103/PhysRevD.104.063524} {\bibfield
  {journal} {\bibinfo  {journal} {Phys. Rev. D}\ }\textbf {\bibinfo {volume}
  {104}},\ \bibinfo {pages} {063524} (\bibinfo {year} {2021})}\BibitemShut
  {NoStop}%
\bibitem [{\citenamefont {Di~Valentino}\ and\ \citenamefont
  {Bridle}(2018)}]{DiValentino:2018gcu}%
  \BibitemOpen
  \bibfield  {author} {\bibinfo {author} {\bibfnamefont {E.}~\bibnamefont
  {Di~Valentino}}\ and\ \bibinfo {author} {\bibfnamefont {S.}~\bibnamefont
  {Bridle}},\ }\href {\doibase 10.3390/sym10110585} {\bibfield  {journal}
  {\bibinfo  {journal} {Symmetry}\ }\textbf {\bibinfo {volume} {10}},\ \bibinfo
  {pages} {585} (\bibinfo {year} {2018})}\BibitemShut {NoStop}%
\bibitem [{\citenamefont {Nunes}\ and\ \citenamefont
  {Vagnozzi}(2021)}]{Nunes:2021ipq}%
  \BibitemOpen
  \bibfield  {author} {\bibinfo {author} {\bibfnamefont {R.~C.}\ \bibnamefont
  {Nunes}}\ and\ \bibinfo {author} {\bibfnamefont {S.}~\bibnamefont
  {Vagnozzi}},\ }\href {\doibase 10.1093/mnras/stab1613} {\bibfield  {journal}
  {\bibinfo  {journal} {Mon. Not. Roy. Astron. Soc.}\ }\textbf {\bibinfo
  {volume} {505}},\ \bibinfo {pages} {5427} (\bibinfo {year} {2021})},\ \Eprint
  {http://arxiv.org/abs/2106.01208} {arXiv:2106.01208 [astro-ph.CO]}
  \BibitemShut {NoStop}%
\bibitem [{\citenamefont {Hikage}\ \emph {et~al.}(2019)\citenamefont {Hikage}
  \emph {et~al.}}]{Hikage_2019}%
  \BibitemOpen
  \bibfield  {author} {\bibinfo {author} {\bibfnamefont {C.}~\bibnamefont
  {Hikage}} \emph {et~al.} (\bibinfo {collaboration} {HSC}),\ }\href {\doibase
  10.1093/pasj/psz010} {\bibfield  {journal} {\bibinfo  {journal} {Publ.
  Astron. Soc. Jap.}\ }\textbf {\bibinfo {volume} {71}},\ \bibinfo {pages} {43}
  (\bibinfo {year} {2019})},\ \Eprint {http://arxiv.org/abs/1809.09148}
  {arXiv:1809.09148 [astro-ph.CO]} \BibitemShut {NoStop}%
\bibitem [{\citenamefont {Asgari}\ \emph {et~al.}(2021)\citenamefont {Asgari}
  \emph {et~al.}}]{Asgari_2021}%
  \BibitemOpen
  \bibfield  {author} {\bibinfo {author} {\bibfnamefont {M.}~\bibnamefont
  {Asgari}} \emph {et~al.} (\bibinfo {collaboration} {KiDS}),\ }\href {\doibase
  10.1051/0004-6361/202039070} {\bibfield  {journal} {\bibinfo  {journal}
  {Astron. Astrophys.}\ }\textbf {\bibinfo {volume} {645}},\ \bibinfo {pages}
  {A104} (\bibinfo {year} {2021})},\ \Eprint {http://arxiv.org/abs/2007.15633}
  {arXiv:2007.15633 [astro-ph.CO]} \BibitemShut {NoStop}%
\bibitem [{\citenamefont {Choi}\ \emph {et~al.}(2020)\citenamefont {Choi},
  \citenamefont {Hasselfield}, \citenamefont {Ho}, \citenamefont {Koopman},
  \citenamefont {Lungu}, \citenamefont {Abitbol}, \citenamefont {Addison},
  \citenamefont {Ade}, \citenamefont {Aiola}, \citenamefont {Alonso},\ and\
  \citenamefont {et~al.}}]{Choi_2020}%
  \BibitemOpen
  \bibfield  {author} {\bibinfo {author} {\bibfnamefont {S.~K.}\ \bibnamefont
  {Choi}}, \bibinfo {author} {\bibfnamefont {M.}~\bibnamefont {Hasselfield}},
  \bibinfo {author} {\bibfnamefont {S.-P.~P.}\ \bibnamefont {Ho}}, \bibinfo
  {author} {\bibfnamefont {B.}~\bibnamefont {Koopman}}, \bibinfo {author}
  {\bibfnamefont {M.}~\bibnamefont {Lungu}}, \bibinfo {author} {\bibfnamefont
  {M.~H.}\ \bibnamefont {Abitbol}}, \bibinfo {author} {\bibfnamefont {G.~E.}\
  \bibnamefont {Addison}}, \bibinfo {author} {\bibfnamefont {P.~A.~R.}\
  \bibnamefont {Ade}}, \bibinfo {author} {\bibfnamefont {S.}~\bibnamefont
  {Aiola}}, \bibinfo {author} {\bibfnamefont {D.}~\bibnamefont {Alonso}}, \
  and\ \bibinfo {author} {\bibnamefont {et~al.}},\ }\href {\doibase
  10.1088/1475-7516/2020/12/045} {\bibfield  {journal} {\bibinfo  {journal}
  {Journal of Cosmology and Astroparticle Physics}\ }\textbf {\bibinfo {volume}
  {2020}},\ \bibinfo {pages} {045–045} (\bibinfo {year} {2020})}\BibitemShut
  {NoStop}%
\bibitem [{\citenamefont {Hill}\ \emph {et~al.}(2022)\citenamefont {Hill} \emph
  {et~al.}}]{Hill_2021_ACT}%
  \BibitemOpen
  \bibfield  {author} {\bibinfo {author} {\bibfnamefont {J.~C.}\ \bibnamefont
  {Hill}} \emph {et~al.},\ }\href {\doibase 10.1103/PhysRevD.105.123536}
  {\bibfield  {journal} {\bibinfo  {journal} {Phys. Rev. D}\ }\textbf {\bibinfo
  {volume} {105}},\ \bibinfo {pages} {123536} (\bibinfo {year} {2022})},\
  \Eprint {http://arxiv.org/abs/2109.04451} {arXiv:2109.04451 [astro-ph.CO]}
  \BibitemShut {NoStop}%
\bibitem [{\citenamefont {Smith}\ \emph {et~al.}(2022)\citenamefont {Smith},
  \citenamefont {Lucca}, \citenamefont {Poulin}, \citenamefont {Abellan},
  \citenamefont {Balkenhol}, \citenamefont {Benabed}, \citenamefont {Galli},\
  and\ \citenamefont {Murgia}}]{Smith:2022hwi}%
  \BibitemOpen
  \bibfield  {author} {\bibinfo {author} {\bibfnamefont {T.~L.}\ \bibnamefont
  {Smith}}, \bibinfo {author} {\bibfnamefont {M.}~\bibnamefont {Lucca}},
  \bibinfo {author} {\bibfnamefont {V.}~\bibnamefont {Poulin}}, \bibinfo
  {author} {\bibfnamefont {G.~F.}\ \bibnamefont {Abellan}}, \bibinfo {author}
  {\bibfnamefont {L.}~\bibnamefont {Balkenhol}}, \bibinfo {author}
  {\bibfnamefont {K.}~\bibnamefont {Benabed}}, \bibinfo {author} {\bibfnamefont
  {S.}~\bibnamefont {Galli}}, \ and\ \bibinfo {author} {\bibfnamefont
  {R.}~\bibnamefont {Murgia}},\ }\href {\doibase 10.1103/PhysRevD.106.043526}
  {\bibfield  {journal} {\bibinfo  {journal} {Phys. Rev. D}\ }\textbf {\bibinfo
  {volume} {106}},\ \bibinfo {pages} {043526} (\bibinfo {year} {2022})},\
  \Eprint {http://arxiv.org/abs/2202.09379} {arXiv:2202.09379 [astro-ph.CO]}
  \BibitemShut {NoStop}%
\bibitem [{\citenamefont {Posta}\ \emph {et~al.}(2021)\citenamefont {Posta},
  \citenamefont {Louis}, \citenamefont {Garrido},\ and\ \citenamefont
  {Hill}}]{LaPosta_2021}%
  \BibitemOpen
  \bibfield  {author} {\bibinfo {author} {\bibfnamefont {A.~L.}\ \bibnamefont
  {Posta}}, \bibinfo {author} {\bibfnamefont {T.}~\bibnamefont {Louis}},
  \bibinfo {author} {\bibfnamefont {X.}~\bibnamefont {Garrido}}, \ and\
  \bibinfo {author} {\bibfnamefont {J.~C.}\ \bibnamefont {Hill}},\ }\href@noop
  {} {\bibfield  {journal} {\bibinfo  {journal} {arXiv e-prints}\ } (\bibinfo
  {year} {2021})},\ \Eprint {http://arxiv.org/abs/2112.10754} {arXiv:2112.10754
  [astro-ph.CO]} \BibitemShut {NoStop}%
\end{thebibliography}%
\bibliographystyle{apsrev4-1}

\end{document}